# Continuous Quantum Aperture: Beamforming with a Single-Vapor-Cell Rydberg Receiver

Mingyao Cui[1,†], Qunsong Zeng[1,†], Minze Chen[2,†], Yilin Wang[2], Zhiao Zhu[2], Tianqi Mao[2,*], Dezhi Zheng[2,*], Kaibin Huang[1,*], Jun Zhang[2]

† Equal contribution
* Corresponding authors (Email: maotq@bit.edu.cn, zhengdezhi@bit.edu.cn, huangkb@hku.hk)
[1] Department of Electrical and Computer Engineering, The University of Hong Kong, Hong Kong SAR, China
[2] The State Key Laboratory of Environment Characteristics and Effects for Near-space, Beijing Institute of Technology, Beijing, China.

**Abstract**

Beamforming is conventionally understood as a collective property of many discrete antenna elements in both communication and radar fields, which links angular selectivity to array size, element spacing, and band-specific hardware. Here we uncover a fundamentally-different beamforming mechanism achieved by a Rydberg atomic receiver: a Rydberg-atom vapor cell dressed by a local-oscillator field constitutes a continuous quantum aperture. In this regime, spatially-varying quantum coherence across the aperture provides continuous amplitude-phase control, allowing a directional beam pattern to emerge from one sensing volume rather than from an engineered array. We establish the theory of continuous quantum aperture and show that tailoring the local-oscillator field can directly program the aperture response. This enables reconfigurable single-peak, multipeak, and multiband beamforming within a single vapor cell. Experiments on a Rydberg atomic receiver prototype verify that practical beam patterns agree with theoretical predictions across aperture sizes, frequency bands, and local-oscillator configurations. Leveraging this new beamforming mechanism, we further demonstrate interference mitigation, multiuser access, and multiband multiuser access with the single-vapor-cell platform. Our results identify the continuous quantum aperture as a new operating principle of Rydberg atomic receivers and establish single-vapor-cell beamforming as an integrated and reconfigurable platform for spatially selective electromagnetic reception.

## Introduction

Beamforming is a foundational operation of modern wireless systems, which enables directional signal transmission and reception for communications, radar, navigation and imaging[1-4]. In conventional implementation, beamforming is realized by combining signals from many discrete antenna elements distributed across a relatively large aperture[5,6]. Despite its superior capability, this architecture is generally constrained by two physical requirements. First, generating high-gain beams requires a massive number of discrete antenna elements[7]. For instance, an extremely large-aperture array for 6G systems can comprise tens of thousands of antennas[8]. Second, to avoid grating lobes and maintain high radiation efficiency, the element spacing should be on the order of half wavelength[9,10]. This geometric requirement ties the hardware to a given spectral range, such that operations across widely separated bands (e.g., S-band, Ku-band, and THz band) have to employ separate antenna arrays[11]. Consequently, these two constraints make integrated and versatile beamforming quite difficult to realize with conventional arrays, motivating the search for a fundamentally different architecture capable of achieving comparable spatial selectivity within a highly-compact medium.

Recent advances in quantum sensing provide such a possibility in the form of the Rydberg atomic receiver[12-15]. Rydberg atoms are atoms in high energy-level states. Owing to their large transition dipole moments, these atoms couple strongly to electromagnetic (EM) fields through two physics phenomena called electromagnetically induced transparency (EIT) and Autler-Townes (AT) splitting[16,17]. This interaction enables high-sensitivity measurement of the amplitude, frequency, phase, and polarization of an incident EM field[18-20]. It also enables a large dynamic range (>120 dB) and system international (SI) traceability to fundamental constants[21,22]. These capabilities have stimulated growing interest in Rydberg atomic receivers for wireless communications and sensing systems. In communications, experiments have demonstrated analogue transmission[23-26] (e.g., amplitude/frequency/phase modulation), digital modulations[27,28] (e.g., quadrature amplitude modulation (QAM)), multifrequency detection[29,30], anti-jamming communications[31], and multiple-input-multiple-output communications[32,33]. In sensing, Rydberg atomic receivers have enabled displacement detection[34], multiband sensing[35,36], angle-of-arrival estimation[36,37], radar ranging[38,39], moisture sensing[40], and discharge recognition[41]. Beyond high sensitivity, Rydberg atomic receivers offer two appealing attributes for beamforming. First, because the inter-atomic spacing is on the order of micrometers[18,26,42], a Rydberg-atom vapor cell can be regarded as a continuous receiving aperture comprising an infinite number of "quantum antennas". Second, the abundant energy levels of Rydberg atoms enable a single vapor cell to sense EM fields across an ultra-wide spectrum spanning from MHz to THz[43,44]. These advantages suggest a possible paradigm shift in beamforming towards using a single atomic vapor cell as a continuous, ultra-wideband receiving aperture.

Despite its promising potential, the intrinsic beamforming capability of a single vapor cell has remained unexplored, because such cells have long been regarded as omnidirectional antennas[12,17,45]. Specifically, under *EIT-AT detection scheme*, the optical response is merely determined by the incident field amplitude but independent of its direction of arrival, which thereby yields an angle-invariant EIT-AT spectrum. This omnidirectional response has been established both theoretically and experimentally in the literature [45]. In this paper, however, we reveal that such omnidirectional-response picture breaks down under the more sensitive and popular *superheterodyne detection scheme* [26,29,30]. In this scheme, an external local-oscillator (LO) field is introduced to dress the Rydberg atom states, which can substantially enhance detection sensitivity and has become a standard configuration for Rydberg atomic receivers. We discover that such an LO-dressed receiver structure exhibits a remarkable and controllable angular response pattern. Based on this finding, we introduce the concept of continuous quantum aperture, in which a continuous atomic medium (a single LO-dressed vapor cell) can perform efficient and reconfigurable beamforming.

Specifically, we establish the operating principle of quantum-aperture beamforming. It is shown that, under LO-dressing, spatially-varying quantum coherence is formed across the Rydberg-atom medium. Such coherence is mathematically equivalent to a continuous-aperture integral of the phase-rotated incident signal field, with the LO field functioning as a virtual continuous phased array. Consequently, the receiver, even with a single vapor cell, forms a beam pointing towards the opposite direction of the LO source, and its beamwidth narrows as the LO frequency and cell aperture increase. Building on this principle, we further develop a set of advanced quantum beamforming techniques, including multipeak beamforming using multi-LO dressing and multiband beamforming via multiband-LO dressing. To validate our theory, we build a Rydberg atomic receiver prototype and measure over 100 beamforming patterns across varying vapor-cell apertures (4-10 cm), frequency bands (S-band and Ku-band), and beam shapes (single-peak and double-peak). All measurements agree well with our theoretical predictions, validating the theory of quantum-aperture beamforming. Last, we demonstrate its practical value in real-world communications, including interference mitigation, multiuser access, and multiband multiuser access. In interference mitigation experiments, increasing the cell aperture from 4 cm to 10 cm suppresses external interference by 10 dB and reduces the bit error rate (BER) by orders of magnitude. In multiuser access experiments, double-peak beamforming supports two-user access simultaneously, with uncoded BERs below $10^{-3}$ for both users. In multiband multiuser access, multiband beamforming concurrently serves heterogeneous devices operating at S-band and Ku-band, achieving uncoded BERs below $10^{-4}$ in both bands. These results establish the continuous quantum aperture as an integrated and versatile beamforming platform for next-generation wireless systems.

## Results

### Operating principles of quantum-aperture beamforming

**Single-peak Beamforming.** We first describe the operating principles of quantum-aperture beamforming with a single peak, which originates from the spatially-varying quantum coherence of Rydberg atoms. As illustrated in Fig. 1a, the core of our system is a cylindrical vapor cell filled with alkali atoms (Cesium-133), which serves as a *continuous quantum aperture* of length $L$. We excite the atoms via two-photon excitation: a *probe laser* drives the transition from the ground state $|1\rangle$ to the intermediate state $|2\rangle$, and a *coupling laser* further excites the atoms to the Rydberg state $|3\rangle$ (Fig. 1b). To perform superheterodyne detection, an external LO field is introduced to illuminate the aperture from a horizontal direction $\theta_l$, which triggers the Rydberg transition $|3\rangle \rightarrow |4\rangle$ (see Hamiltonian operator in Methods). This LO-dressed quantum aperture detects a signal (SIG) field arriving from direction $\theta_s$. For implementation convenience, the LO and SIG impinge from opposite sides ($\theta_s \in [0, 180°]$ and $\theta_l \in [180°, 360°]$), while our discussion applies equally to co-incident configurations.

At position $y \in [0, L]$ along the aperture, the LO and SIG fields are modelled as $E_l(y,t) = E_l e^{j(\omega_l t + k_l y \cos\theta_l + \phi_l)}$ and $E_s(y,t) = E_s e^{j(\omega_s t + k_s y \cos\theta_s + \phi_s)}$, where $E_l$ and $E_s$ denote the field amplitudes, $\omega_l$ and $\omega_s$ the carrier angular frequencies, $k_l$ and $k_s$ the wavenumbers, and $\phi_l$ and $\phi_s$ the initial phases. Throughout the paper, the subscripts $l$ and $s$ denote the LO and SIG fields, respectively. The frequency offset $\omega_\delta = (\omega_s - \omega_l)$ defines the system's intermediate frequency (IF). Because the two fields arrive from different directions, their phase profiles vary differently along the spatial coordinate. Their interference inside the vapor cell thus jointly modulates the Rydberg states, which results in a *spatially varying quantum coherence*, $\bar{\rho}_{21}$, across the aperture (Fig. 1c). Such variation imposes a position-dependent absorption of the atomic medium to the probe laser as it propagates through the vapor cell. As a result, the cumulative absorption yields an alternating-current (AC) component, $\Delta P(t)$, in the transmitted probe-laser power formulated as (see proof in Methods):

$$\Delta P(t) \propto \text{Re}\left\{\int_0^L \frac{\partial \text{Im}[\bar{\rho}_{21}]}{\partial E_l} e^{-j\angle E_l(y,t)} E_s(y,t) dy\right\} \propto \frac{\partial \text{Im}[\bar{\rho}_{21}]}{\partial E_l} \text{sinc}\left[\frac{L}{\lambda_l}(\cos\theta_s - \cos\theta_l)\right] \cdot E_s \cos(\omega_\delta t + \phi'), \quad (1)$$

which gives rise to a beamforming pattern

$$F_{\text{sp}}(\theta_s) = \frac{\Delta P(t)^2}{\max \Delta P(t)^2} = \text{sinc}^2\left[\frac{L}{\lambda_l}(\cos\theta_s - \cos\theta_l)\right], \qquad (2)$$

with $\text{sinc}(x) = \sin(\pi x)/(\pi x)$ and $\lambda_l = 2\pi c/\omega_l$ the LO wavelength.

Eq. (1) reveals that, at each position $y$, the signal field, $E_s(y,t)$, is rotated by the LO phase, $\angle E_l(y,t)$. The resulting AC laser, $\Delta P(t)$, therefore represents the cumulative contribution of all locally phase-rotated signal fields along the aperture. In this sense, the LO phase profile fundamentally functions as a *virtual continuous phased array* that determines the beamforming pattern, $F_{\text{sp}}(\theta_s)$. When the signal arrives from the direction opposite to the LO ($\theta_s = 360° - \theta_l$), the AC response is maximized; while as $\theta_s$ deviates from $(360° - \theta_l)$, the response is suppressed by the pattern $F_{\text{sp}}(\theta_s)$ due to LO-SIG phase-profile mismatch. As a result, this quantum aperture forms a beam pointing to $(360° - \theta_l)$, and beam steering is achievable by rotating the LO field (Fig. 1a). The half-power beamwidth (HPBW), which quantifies the directivity, is calculated as: $\theta_{\text{HPBW}} = 0.886\frac{\lambda_l}{L}$ [Rad]. Thus, a longer aperture, $L$, and a shorter wavelength, $\lambda_l$, produce a narrower beam. Notably, this beamforming functionality is achieved with a *single* vapor cell: the near-zero inter-atomic spacing effectively provides a continuous array with *an infinite number of "quantum antennas"*, eliminating the bulky and costly radio-frequency (RF) front-end required by a conventional large antenna array.

**Multipeak beamforming.** Beyond single-peak beamforming, the virtual continuous-array nature of the quantum aperture enables more general beam patterns by deliberately programming the LO phase profile. Particularly, multipeak beamforming can be realised by illuminating the cell with multiple LO sources, say $N$ sources, from distinct directions (see Fig. 1d). In this case, the total LO field generalizes to $E_l(y,t) = \sum_{n=1}^{N} E_{l,n} e^{j(\omega_l t + k_l y \cos\theta_{l,n} + \phi_{l,n})} = E_l(y) e^{j\omega_l t + j\phi_l(y)}$, where $E_l(y) = \left|\sum_{n=1}^{N} E_{l,n} e^{jk_l y \cos\theta_{l,n} + j\phi_{l,n}}\right|$ is the multi-LO field amplitude, $\phi_l(y) = \angle\left(\sum_{n=1}^{N} E_{l,n} e^{jk_l y \cos\theta_{l,n} + j\phi_{l,n}}\right)$ the multi-LO phase profile, and $E_{l,n}$, $\theta_{l,n}$, and $\phi_{l,n}$ are the amplitude, incident angle, and initial phase of the $n$-th LO. The AC probe laser of this multi-LO-dressed quantum aperture exhibits a multipeak beamforming pattern (see proof in Methods):

$$F_{\text{mp}}(\theta_s) \propto \left|\int_0^L \frac{\partial \text{Im}(\bar{\rho}_{12})}{\partial E_l(y)} e^{j[k_s y \cos\theta_s - \phi_l(y)]} \mathrm{d}y\right|^2. \qquad (3)$$

In contrast to the single-LO case, the beam pattern $F_{\text{mp}}(\theta_s)$ now depends on the full parameter set $\{\theta_{l,n}\}$, $\{E_{l,n}\}$, and $\{\phi_{l,n}\}$. The LO directions $\{\theta_{l,n}\}$ primarily set the angular locations where Rydberg atoms exhibit strong responses, thereby producing multiple beam peaks. The amplitudes, $\{E_{l,n}\}$, control each LO's contribution to the overall phase profile $\phi_l(y)$: a stronger amplitude, $E_{l,n}$, gives LO $n$ greater contribution, which raises the beam peak near $\theta_s = (360° - \theta_{l,n})$. Accordingly, by jointly adjusting the directions and powers of the LO sources, the beam shape, including peak locations and relative heights, can be flexibly configured. The spatial integral in Eq. (3) is additionally weighted by $\frac{\partial \text{Im}(\bar{\rho}_{12})}{\partial E_l(y)}$. This factor arises because multi-LO dressing produces a spatially nonuniform LO-field strength along the cell, making the sensitivity of the Rydberg atoms position-dependent. This effect is captured precisely by the derivative term, $\frac{\partial \text{Im}(\bar{\rho}_{12})}{\partial E_l(y)}$, as it reflects the sensitivity of quantum coherence to weak perturbations.

**Multiband beamforming.** A quantum aperture can also support simultaneous beamforming across multiple, widely separated frequency bands, referred to as multiband beamforming. This is achieved by exploiting multi-level Rydberg transitions. As shown in Fig. 1e, $N$ LO sources at distinct frequency bands, $\{\omega_{l,n}\}$, and incident directions, $\{\theta_{l,n}\}$, produce a composite field, $E_l(y,t) = \sum_{n=1}^{N} E_{l,n} e^{j(\omega_{l,n} t + k_{l,n} y \cos\theta_{l,n} + \phi_{l,n})}$. These LO tones drive distinct Rydberg transitions $|3\rangle \rightarrow$

$|n+3\rangle, \forall n \in \{1, \cdots, N\}$, (e.g., 3.39 GHz for $|59D_{5/2}\rangle \to |60P_{3/2}\rangle$ and 15.59 GHz for $|59D_{5/2}\rangle \to |57F_{7/2}\rangle$). As derived in Supplementary Note 1, the AC probe laser of this multiband-LO-dressed quantum aperture is

$$\Delta P(t) \propto \sum_{n=1}^{N} \alpha_n \frac{\partial \text{Im}[\bar{\rho}_{21}]}{\partial E_{l,n}} \text{sinc}\left[\frac{L}{\lambda_{l,n}}(\cos\theta_{s,n} - \cos\theta_{l,n})\right] \cdot E_{s,n} \cos(\omega_{\delta,n} t + \phi'_n). \quad (4)$$

where $\lambda_{l,n} = 2\pi c/\omega_{l,n}$, $E_{s,n}$, $\omega_{\delta,n}$, and $\theta_{s,n}$ are the wavelength, amplitude, IF, and SIG field direction at band $n$, and $\alpha_n$ is a constant.

Eq. (4) shows that in each band, an independent beam is generated with a unique beam pattern $F_{\text{sp},n}(\theta_{s,n}) = \text{sinc}^2\left[\frac{L}{\lambda_{l,n}}(\cos\theta_{s,n} - \cos\theta_{l,n})\right]$. The beam is steered toward the direction opposite to the corresponding LO, ($\theta_{s,n} = 360° - \theta_{l,n}$) and the beamwidth scales as $0.886\frac{\lambda_{l,n}}{L}$ [Rad]. Notably, conventional multiband beamforming typically requires dedicated antenna arrays and RF front ends for each band: higher frequencies often require more antenna elements to preserve a fixed physical aperture. By contrast, Eq. (4) dictates that, leveraging the continuous-array nature and the abundant Rydberg energy levels, a single quantum aperture is sufficient to provide beamforming across widely separated bands without modifying its physical configuration. This enables a unified platform that offers beamforming gain to heterogeneous wireless devices, like a mobile user in the S-band and a satellite in the Ku-band.

**Experimental validation of quantum-aperture beamforming**

We implemented a Rydberg atomic receiver prototype (Fig. 2a) to experimentally demonstrate quantum-aperture beamforming. To quantify the impact of cell aperture on directivity, we prepared vapor cells with lengths from 4 cm to 10 cm (Fig. 2b). To suppress multipath scattering, the atomic vapor cell was housed in a semi-cylindrical anechoic chamber. Cesium-133 atoms were excited to the Rydberg state $|59D_{5/2}\rangle$ using an 852-nm probe laser and a 509-nm coupling laser. A window was opened at the back of the chamber to permit the propagation of the LO field. In front of the chamber, we mounted a 1.4-m-diameter semicircular rail centred on the vapor cell. The SIG horn antenna was attached to the rail. This configuration allows precise rotation of the SIG incidence direction, and thereby the accurate measurement of beamforming pattern (see Methods for detailed setup). For comparison, we measured the cell's angular response using two detection schemes: (i) EIT-AT detection without an LO[45] and (ii) superheterodyne detection with LOs. As illustrated in Fig. 2c, EIT-AT detection estimates the SIG strength from the measured AT splitting via two-Lorentzian peak fitting (see Supplementary Note 4). In contrast, superheterodyne detection, the focus of LO-dressed quantum aperture, locks the laser frequencies and records the AC amplitude of the transmitted probe power at a fixed intermediate frequency ($\omega_\delta = 2\pi \times 5\text{kHz}$), which is expected to follow the derived beamforming patterns.

We first validate single-peak beamforming and multiband beamforming. Experiments were carried out at two separated frequency bands: 3.39 GHz (S-band, $|59D_{5/2}\rangle \to |60P_{3/2}\rangle$) and 15.59 GHz (Ku-band, $|59D_{5/2}\rangle \to |57F_{7/2}\rangle$). Three LO directions were tested: $\theta_l = 300°, 270°$ and $240°$, which correspond to three beam directions $(360° - \theta_l) = 60°, 90°$ and $120°$, respectively. For each trial, the SIG source was swept from 20° to 160° along the rail to measure the beam pattern. The angular step size was set to 5° for EIT-AT detection, 5° for superheterodyne detection at 3.39 GHz, and 2° for the heterodyne detection at 15.59 GHz. The measured beam patterns and extracted HPBWs are presented in Fig. 2d–f. Across both frequency bands and all cell apertures, the measured single-peak patterns $\bar{F}_{sp}(\theta_s)$ (solid lines) closely agree with theoretical predictions $F_{sp}(\theta_s)$ (dashed lines) from Eq. (2) and (4). The beams are accurately steered toward the directions opposite to the LO incidence (i.e., $\theta_s = 60°, 90°$ and $120°$), confirming the beam steering capability. Moreover, the measured HPBW narrows systematically with increasing frequency and cell aperture $L$, in quantitative agreement with the theoretical scaling $0.886\lambda_l/L$. The root-mean-square (RMS) beamwidth errors between experiment and theory are 11.80° at 3.39 GHz and 2.91° at 15.59 GHz. Additional beam patterns measured at 3.77GHz ($|57D_{5/2}\rangle \to |58P_{3/2}\rangle$) and 12.71GHz ($|63D_{5/2}\rangle \to |61F_{7/2}\rangle$), associated with their estimated beamwidths are provided in Supplementary Notes 5-6. They all show the same agreement with theoretical predictions, validating the established theory for quantum-aperture beamforming.

For comparison, the angular response patterns obtained via EIT-AT detection are shown as shaded regions in Fig. 2e, f. At 3.39 GHz, the EIT-AT pattern is nearly omnidirectional, in remarkable contrast to the directional pattern obtained via superheterodyne detection. At 15.59 GHz, the EIT-AT pattern exhibits distortions that we attribute to scattering near the corners of the cylindrical vapor cell for shorter-wavelength signals. These distortions cause localized attenuation and produce apparent valleys in the measured pattern. Similar distortions have also been reported previously in the study[45]. In Fig. 2f, these valleys occur near 60° and 120°, indicating reduced incident power at those angles. Nevertheless, superheterodyne detection still produces sharp beam peaks at 60° and 120° that effectively suppress the signal attenuation caused by cell-corner scattering. This comparison demonstrates that the beam directivity originating from spatially-varying quantum coherence dominates over pattern distortions caused by the geometry of vapor cells.

We next validate double-peak beamforming. In this experiment, two LO horn antennas radiated 15.59 GHz tones from different directions to an 8-cm vapor cell. Eq. (3) predicts that such a double-LO-dressed quantum aperture should generate two beam peaks pointing opposite to the LO directions. The experimental results, shown in Fig. 2g, confirm this prediction, where the left three panels of Fig. 2g correspond to LO incidence directions at (270°, 240°) and the right three to (300°, 240°). For each configuration, we tested three LO power allocations: (LO1, LO2) = (3 dBm, 6 dBm), (4 dBm, 4 dBm), and (6 dBm, 3 dBm). Solid curves show the measured patterns $\bar{F}_{\mathrm{mp}}(\theta_s)$, while dashed curves denote the theoretical fittings $F_{\mathrm{mp}}(\theta_s)$ from Eq. (3). Because the initial LO phases inside the vapor cell, $\phi_{l,1}$ and $\phi_{l,2}$, are not directly measurable, we treated them as fitting parameters and minimized the residual $\sum_{\{\phi_{l,n}\}} |\bar{F}_{\mathrm{mp}}(\theta_s) - F_{\mathrm{mp}}(\theta_s)|^2$. The resulting fits show close agreement with experimental results. The beam peaks are accurately steered to (90°, 120°) and (60°, 120°), respectively. In addition, both theory and experiment confirm that the relative peak heights can be tuned by adjusting the LO power allocation. An equal LO-power distribution can produce peaks of comparable height (the 2[nd] and 5[th] panels in Fig. 2g); when LO1 is weaker than LO2, a left-low, right-high beam pattern emerges (the 1[st] and 4[th] panels in Fig. 2g); the converse yields a left-high, right-low pattern (the 3[rd] and 6[th] panels in Fig. 2g). Together, these results validate our theoretical predictions and demonstrate programmable shaping of beamforming patterns to match different communication requirements.

**Quantum-aperture beamforming enabled wireless communications**

**Interference mitigation.** The preceding discussion demonstrates that a continuous quantum aperture inherently performs beamforming. Such spatial selectivity is particularly useful for *interference mitigation*, a key application in real-world communication systems. The high sensitivity of Rydberg atomic receivers, while advantageous for signal detection, also leaves them susceptible to external interferers. By acting as a spatial filter, the quantum aperture enables the suppression of these unwanted signals.

To illustrate this capability, we implemented a real-world communication scenario in which the quantum aperture discriminates between a desired user and an interferer (Fig. 3a). The LO was aligned so that the vapor cell formed a beam directed toward the target user. The user transmitted 16-QAM symbols at frequency 15.59 GHz, with a symbol rate of 4 kSymbols/s and a received power spectral density of -50 dBV/Hz. Simultaneously, a strong interferer, positioned at an angular separation of 15° from the desired source, emitted random signals with a bandwidth of 8 kHz. We controlled the power of the interferer to achieve a range of transmit signal-to-interference ratios (SIRs) from -6 dB to 6 dB. Moreover, the length of the vapor cell was increased from 4 cm to 10 cm to quantify the impact of cell aperture on the interference-mitigation capability. For each length and SIR, the receiver demodulated the user symbols using the method in Supplementary Note 3, and we recorded the receive spectrum, constructed constellation diagrams, and computed the error-vector magnitude (EVM) and BER.

The dependence of EVM and BER on both cell length and SIR is presented in Fig. 3b, c. In all cases, EVM and BER decrease monotonically as either the cell length or SIR increases. Quantitatively, EVM decreases from 30.86% to 9.03% and BER declines from $1.68 \times 10^{-1}$ to $7.62 \times 10^{-5}$ over the tested range. This improvement stems from the progressive narrowing of the reception beam with longer cell lengths. The resulting beam-narrowing suppresses the interferer, lowering its received power spectral density from approximately $-55$ dBV/Hz to about $-65$ dBV/Hz. This suppression effectively raises the operational SIR at the receiver and thereby enhances the system's interference-mitigation capability. The corresponding enhancement in signal integrity is visually corroborated by the constellation diagrams in Fig. 3d, which depict 16-QAM results under a representative SIR of 0 dB. As the cell length increases from 4 cm to 10 cm, the symbol clusters transition from widely scattered and distorted distributions to tightly grouped, well-defined points. For comparison, system performance was also evaluated under Quadrature Phase Shift Keying (QPSK) modulation (see Supplementary Note 7). All the aforementioned trends, including beam narrowing, SIR enhancement, and the dependence of EVM and BER on cell length, apply similarly to QPSK. These results verify the interference-mitigation capability of quantum-aperture beamforming, underscoring its potential for deployment in strong interference environments.

**Multiuser access.** Another key application of quantum-aperture beamforming is multiuser access. In this scenario, we employed two LOs to generate a double-peak beam. Each peak can enhance signal reception in a specific direction, thereby establishing the spatially selective communication links with two users (Fig. 4a). As detailed previously, the relative intensities of the two LO fields directly control the heights of these two peaks. This, in turn, enables dynamic allocation of channel gain and allows for the real-time tuning of communication performance for each user link.

In the implementation, the two LOs were positioned at fixed directions of 300° (LO1) and 270° (LO2), both transmitting at 15.59 GHz. The cell length was 8 cm. Two users, denoted by UE1 and UE2, were each aligned with one of the generated beam peaks. UE1 transmitted an image of The University of Hong Kong (HKU) logo, modulated using 16-QAM onto a 24 kHz IF at a symbol rate of 4 kSymbols/s. Simultaneously, UE2 transmitted an image of the Beijing Institute of Technology (BIT) logo, modulated using identical 16-QAM parameters, but onto a 33 kHz IF, also at a rate of 4 kSymbols/s. With all modulation and encoding parameters held constant, we varied the transmit power of the two LOs. Specifically, the power of LO1 was increased stepwise from 6 dBm to 10 dBm, while the power of LO2 was correspondingly decreased from 6 dBm to 2 dBm. This power sweep directly modulated the relative heights of the two beam peaks, and the resulting communication performance for each user link was evaluated at each power level.

Experimental results align with theoretical expectations. As the power difference between the two LOs increases, the received signal spectrum for UE1 progressively strengthens (Fig. 4d). The corresponding EVM declines from 21.9% to 14.5% (Fig. 4b), and BER decreases from 2.9% to 0.3% (Fig. 4c), indicating an improved communication quality. Accordingly, the received HKU logo transitions from blurry to clearly defined (Fig. 4e). In contrast, the received signal spectrum for UE2 weakens (Fig. 4d), with EVM increasing from 11.9% to 25.4% (Fig. 4b) and BER rising from 0.1% to 4.0% (Fig. 4c); the BIT logo deteriorates from clear to indistinct (Fig. 4e). To further validate this approach, we reconfigured the LOs' spatial arrangement, fixing one at 300° and relocating the other to 240°. This modified arrangement reproduced the same beamforming and dynamic power allocation behaviour (see Supplementary Fig. 6). These results demonstrate that by controlling the relative intensities of the dual beams, communication performance can be dynamically adapted to meet the varying requirements of different users with different tasks.

**Multiband multiuser access.** Beyond conventional single-band multiuser access, quantum-aperture beamforming can naturally support multiband multiuser access, which is difficult to achieve for classical antenna arrays. As discussed earlier, leveraging multiband-LO dressing, the quantum aperture can provide beamforming service at largely separated frequency bands. This property therefore allows for concurrent uplink access for heterogeneous devices operating at different bands, for example, a S-band mobile user and a Ku-band satellite terminal.

We implemented a real-world communication scenario in Fig. 5a to validate this application. Two LO horns, operating at 3.39 GHz and 15.59 GHz illuminated the vapor cell from directions 300° and 240°, respectively. The per-band beams were steered toward an S-band signal source (SIG1) at 60° and a Ku-band signal source (SIG2) at 90°. IFs were set to 24 kHz for SIG1 and 33 kHz for SIG2. Each signal source transmitted 16-QAM symbols at a rate of 4 kSymbols/s. To examine the effect of beam directivity, we deliberately misaligned each signal source from its beam centre: SIG1 was deviated by $0 \sim 40°$ and SIG2 by $0 \sim 20°$, while communication performance was recorded at each offset.

Take the 8-cm vapor cell as an example. As shown in Fig. 5 b,c,d, the communication performance is the best for perfect alignment (0° misalignment). The received signal spectra exceed -60 dBV/Hz and the recovered constellations are well resolved (Fig. 5d), yielding a BER of $3.7 \times 10^{-5}$ with an EVM of 8.34% for SIG1, and a BER of $2.1 \times 10^{-4}$ with an EVM of 10.59% for SIG2. Beam misalignment degrades the communication performance differently for the two bands due to their distinct beamwidths. For the Ku-band with a narrow beam, a 20° offset reduces the signal spectrum by 15 dB (Fig. 5f), increases the BER to $1.1 \times 10^{-1}$, and raises the EVM to 33.4 %. In contrast, the S-band link, with a wide beam, shows negligible degradation at 20° misalignment; even at 40° offset, the spectrum drops by only 5 dB (Fig. 5e), the BER rises to $2.8 \times 10^{-3}$, and the EVM to 16.2 %. These results indicate that, in practical communication systems, quantum-aperture beamforming at higher frequencies requires more precise beam alignment to maintain reliable performance. The same trend is observed for a 5-cm vapor cell (Fig. 5b,c) and for LO directions of 240° and 270° (see Supplementary Note 9).

## Discussion

In summary, we have shown that an LO-dressed Rydberg-atom vapor cell can function as a continuous quantum aperture. By exploiting spatially varying quantum coherence within the cell, it enables quantum-aperture beamforming with an effectively infinite number of quantum antennas and seamless operations across widely separated frequency bands. We have built a quantum-aperture beamforming prototype to validate that diverse beam patterns, including single-peak, multipeak, and multiband responses, can be generated by programming the LO field. We further demonstrated several application-relevant scenarios. First, quantum-aperture beamforming was shown as a practical route to interference mitigation. Specifically, although Rydberg atomic receivers are highly sensitive and therefore susceptible to external interferers, the long quantum aperture provides high spatial selectivity to suppress unwanted signals arriving from other directions. Second, multipeak beamforming was implemented to support multiuser access and enable flexible, per-user performance tuning via control of the relative beam-peak heights. Third, we proposed and experimentally verified multiband multiuser access, in which a continuous quantum aperture provides beamforming gain to heterogeneous devices operating in distinct frequency bands.

The proposed continuous quantum aperture offers several advantages over conventional antenna arrays. First, it overcomes the classical half-wavelength sampling constraint, which thus enables higher spatial resolution and efficient beamforming across arbitrary frequency bands. By contrast, achieving comparable functionality with conventional systems would require hardware-intensive, band-specific antenna elements and RF front-end circuitry. In addition, classical engineered arrays must undergo careful amplitude/phase/timing/position calibration for each antenna element to ensure consistency across multiple receive modules. These complex calibrations are unnecessary in a quantum aperture, because its beamforming capability arises directly from the quantum-coherence distribution of Rydberg atoms and all of its quantum antennas share a common receive module.

More broadly, our proposed theory and design substantially expand the application scope of Rydberg atomic receivers, with potential impact on wireless communications, radar sensing, holographic imaging, and related areas. Beyond the presented demonstrations, quantum-aperture beamforming can also facilitate single-vapor-cell angle-of-arrival sensing. This is achieved by sweeping the LO field and identifying the maximum response. Future works can also explore digitally programmable LO field for real-time beam management. A potential approach is to replace the LO horn antenna by a leaky-wave antenna comprising multiple radiating slots controlled by field-programmable gate array (FPGA).

## Methods

**Experiment setup.** On the transmitter side, we employed Ceyear 1466H-V and Keysight E8267D as the two SIG sources, which can generate signals covering S-band and Ku-band. For beam pattern measurement experiments, only the Ceyear 1466H-V was used to send a single tone. For communication experiments, both generators were used to send 16-QAM symbols. All symbols were pulse-shaped using a square-root-raised-cosine (SRRC) filter with a roll-off factor of 0.35 and were then emitted through horn antennas (LB-20180-SF) directed at the vapor cell.

On the receiver side, we have prepared alkali atom (Cesium-133) vapor cells with apertures ranging from 4 cm to 10 cm. Two PXIE 5654 devices served as LO sources, and two horn antennas (LB-20180-SF) connected to the LO sources were positioned behind the anechoic chamber to emit LO fields. The atoms were excited from the ground state $|6S_{1/2}\rangle$ to an intermediate state $|6P_{3/2}\rangle$ by an 852nm probe laser (1.5mW), and from $|6P_{3/2}\rangle$ to the Rydberg state $|59D_{5/2}\rangle$ by a 509nm coupling laser (40mW). The probe and coupling lasers were generated by CoSF-852nm and CoSF-509nm, respectively. For experimental convenience, their optical paths were arranged on opposite sides of the vapor cell with a separation of 1.2 m. The frequency of the 852 nm probe laser was locked using a saturated-absorption spectroscopy (SAS) module, while its power was stabilized by an electronic variable optical attenuator (V800PA) combined with a servo loop. The frequency of the 509 nm coupling laser was locked using an additional EIT signal, and its power was stabilized by an acousto-optic modulator (AOM, Gooch & Housego 3307) combined with a servo loop. The probe laser was split by a beam displacer (Thorlabs BD40) into two parallel beams with 4 mm separation, while the vertically polarized component was overlapped with the coupling laser. After passing through the vapor cell, the two probe beams were received by a balanced photodetector (Hamamatsu S16008-33) to perform differential amplification, thereby suppressing common-mode background noise. The resulting photocurrent was further amplified by a Transimpedance amplifier, and finally recorded by a digital oscilloscope (MSO56B, Tektronix). More illustrations on the experimental system can be found in the Supplementary Note 2.

**Lindblad master equation.** For a four-level ladder-type quantum system, the Hamiltonian in the interaction picture at position $y \in [0, L]$ along the cell is given by

$$H(y,t) = \frac{\hbar}{2}\begin{bmatrix} 0 & \Omega_p & 0 & 0 \\ \Omega_p & -2\Delta_p & \Omega_p & 0 \\ 0 & \Omega_c & -2\Delta_p - 2\Delta_c & \Omega_l + \Omega_s e^{jS(y,t)} \\ 0 & 0 & \Omega_l + \Omega_s e^{-jS(y,t)} & -2\Delta_p - 2\Delta_c - 2\Delta_l \end{bmatrix}, \tag{5}$$

Here, $\Delta_{p,c,l}$ represent the detunings of the probe, coupling, and LO fields, respectively, and $\Omega_{p,c}$ are the probe and coupling Rabi frequencies. We define $\Omega(y,t) = \Omega_l + \Omega_s e^{jS(y,t)}$ as the Rabi frequency of the composite field, where $\Omega_l = \frac{\mu_{34}}{\hbar} E_l$ and $\Omega_s = \frac{\mu_{34}}{\hbar} E_s$ represent the Rabi frequencies of the LO and SIG fields, respectively, $\mu_{34}$ denotes the transition dipole moment, and $S(y,t) = (\angle E_s(y,t) - \angle E_l(y,t))$ characterizes the LO-SIG phase difference. The dynamics of the four-level density matrix $\rho$ is governed by the Lindblad master equation

$$\dot{\rho} = -\frac{j}{\hbar}[H(y,t), \rho] + \mathcal{L}[\rho], \tag{6}$$

where $[H, \rho] = H\rho - \rho H$ is the commutator. The decoherence operator is $\mathcal{L}[\rho] = -\frac{1}{2}\Gamma\rho - \frac{1}{2}\rho\Gamma + \Lambda$, where $\Gamma = \text{diag}\{0, \gamma_2, \gamma_3, \gamma_4\}$, $\Lambda = \text{diag}\{\gamma_2\rho_{22} + \gamma_4\rho_{44}, \gamma_3\rho_{33}, 0, 0\}$, and $\gamma_i$ ($i = 2,3,4$) are the spontaneous decay rates from the $i$-th level. The atomic medium's susceptibility to the probe laser is determined by the instantaneous steady-state of the quantum coherence $\rho_{21}$, i.e., the (2,1)-th element of $\rho$. In general, $\rho_{21}$ depends on both $\Omega(y,t) = \Omega_l + \Omega_s e^{jS(y,t)}$ and its complex conjugate: $\rho_{21} = \rho_{21}(\Omega, \Omega^*) = \rho_{21}(\Omega_l + \Omega_s e^{jS(y,t)}, \Omega_l + \Omega_s e^{-jS(y,t)})$. Moreover, due to atomic thermal motion, lasers illuminating atoms experience a random Doppler shift. The observed quantum state $\rho_{21}$ should be averaged over all possible shifts as follows

$$\bar{\rho}_{21}(\Omega, \Omega^*) = \int_a^a \frac{e^{-\frac{v^2}{u^2}}}{\sqrt{\pi}u} \rho_{21}(\Omega, \Omega^*)\Big|_{\Delta_p = \Delta_p - \frac{2\pi v}{\lambda_p},\ \Delta_c = \Delta_c + \frac{2\pi v}{\lambda_c}} dv, \tag{7}$$

where $u = \sqrt{2k_B T_{\text{env}}/m}$ with $m$ the atom's mass, $k_B$ the Boltzmann constant, and $T_{\text{env}}$ the environment temperature. Besides, $\frac{2\pi v}{\lambda_p}$ and $\frac{2\pi v}{\lambda_c}$ denote the Doppler shifts of the probe and coupling lasers for atoms at velocity $v$. Clearly, $\bar{\rho}_{21}$ is also a spatially varying function of $\Omega(y,t)$ and $\Omega^*(y,t)$.

**Probe laser transmission.** The spatially varying quantum coherence $\bar{\rho}_{21}$ produces an atomic susceptibility $\chi(y,t) = -\frac{2N_0 \mu_{12}^2}{\epsilon_0 \hbar \Omega_p} \bar{\rho}_{21}$, where $N_0$ is the atomic density, $\mu_{12}$ the transition dipole moment from $|1\rangle$ to $|2\rangle$, and $\epsilon_0$ the vacuum permittivity. As the probe laser propagates through the vapor cell, it thereby undergoes position-dependent attenuation. Applying the Bouguer-Beer-Lambert law, given an input power $P_{\text{in}}$, the probe-laser power at the output of vapor cell is

$$P(t) = P_{\text{in}} \exp\left[-k_p \int_0^L \text{Im}[\chi(y,t)] dy\right], \tag{8}$$

where $k_p$ is the wavenumber of probe laser. In practical implementations, the LO field is configured much stronger than the SIG field $\Omega_l \gg \Omega_s$. Under this strong-LO condition, the susceptibility' imaginary part can be linearized as the superposition of $\Omega_s e^{jS(y,t)}$ and $\Omega_s e^{-jS(y,t)}$:

$$\text{Im}[\chi(y,t)] = \text{Im}[\chi_0(y)] + \frac{\partial \text{Im}[\chi]}{\partial \Omega}\bigg|_{\Omega=\Omega_l} \Omega_s e^{jS(y,t)} + \frac{\partial \text{Im}[\chi]}{\partial \Omega^*}\bigg|_{\Omega=\Omega_l} \Omega_s e^{-jS(y,t)}$$
$$= \text{Im}[\chi_0(y)] + \frac{\partial \text{Im}[\chi]}{\partial \Omega_l}\Omega_s \cos(\angle E_s(y,t) - \angle E_l(y,t)), \qquad (9)$$

where $\frac{\partial \text{Im}[\chi]}{\partial \Omega}\big|_{\Omega=\Omega_l} = \frac{\partial \text{Im}[\chi]}{\partial \Omega^*}\big|_{\Omega=\Omega_l} = \frac{1}{2}\frac{\partial \text{Im}[\chi]}{\partial \Omega_l}$, and $\chi_0(y) = \chi(y,t)|_{\Omega=\Omega_l}$ is the LO-only susceptibility. By substituting Eq. (9) into Eq. (8) and applying the Taylor expansion $e^{-x} = 1 - x + \mathcal{O}(x^2)$, the output probe-laser power is linearized as

$$P(t) = \underbrace{P_{\text{in}} e^{-k_p \int_0^L \text{Im}[\chi_0(y)]dy}}_{\bar{P}} - \underbrace{P_{\text{in}} e^{-k_p \int_0^L \text{Im}[\chi_0(y)]dy} k_p \underbrace{\int_0^L \frac{\partial \text{Im}[\chi]}{\partial \Omega_l}\Omega_s \cos(\angle E_s(y,t) - \angle E_l(y,t)) \, dy}_{\text{spatial integration}}}_{\Delta P(t)} = \bar{P} - \Delta P(t). \quad (10)$$

The decomposed direct-current (DC) component $\bar{P}$ results entirely from the strong LO dressing, whereas the AC component carries the information of the SIG field to be detected. Importantly, the spatial integral in $\Delta P(t)$ arises directly from the spatial variation of the quantum coherence $\bar{\rho}_{12}$ along the aperture.

**Derivation of single-peak beamforming.** We now derive the single-peak beamforming phenomenon of a quantum aperture illuminated by a single LO. By substituting the expression of LO and SIG fields into the Rabi frequency $\Omega(y,t) = \Omega_l + \Omega_s e^{jS(y,t)}$, we obtain $\Omega_l = \frac{\mu_{34}}{\hbar}E_l$, $\Omega_s = \frac{\mu_{34}}{\hbar}E_s$, and the LO–SIG phase difference

$$S(y,t) = \angle E_s(y,t) - \angle E_l(y,t) = (\omega_s t + k_s y \cos\theta_s + \phi_s) - (\omega_l t + k_l y \cos\theta_l + \phi_l)$$
$$= \omega_\delta t + (k_s \cos\theta_s - k_l \cos\theta_l)y + \phi_\delta \approx \omega_\delta t + k_l(\cos\theta_s - \cos\theta_l)y + \phi_\delta, \qquad (11)$$

where $\phi_\delta = \phi_s - \phi_l$, and the approximation holds because $k_s \approx k_l$. Further substituting Eq. (11) into Eq. (10) yields the analytical expression of the laser AC component

$$\Delta P(t) = P_{\text{in}} e^{-k_p \int_0^L \text{Im}[\chi_0(y)]dy} k_p \frac{\partial \text{Im}[\chi]}{\partial \Omega_l}\Omega_s \int_0^L \cos(\omega_\delta t + k_l(\cos\theta_s - \cos\theta_l)y + \phi_\delta) \, dy,$$
$$= P_{\text{in}} e^{-k_p \int_0^L \text{Im}[\chi_0(y)]dy} k_p \frac{\partial \text{Im}[\chi]}{\partial \Omega_l}\text{sinc}\left[\frac{L}{\lambda_l}(\cos\theta_s - \cos\theta_l)\right]\Omega_s \cos(\omega_\delta t + \phi'_\delta), \qquad (12)$$

where $\phi'_\delta = \phi_\delta - \pi L(\cos\theta_s - \cos\theta_l)/\lambda_l$. Here, $\frac{\partial \text{Im}[\chi]}{\partial \Omega_l}\Omega_s$ is taken outside the integral because the LO and SIG Rabi frequencies $\Omega_l$ and $\Omega_s$ are spatially-uniform under the plane-wave illumination. Eq. (12) reveals that the time-oscillating signal current $\Omega_s \cos(\omega_\delta t + \phi'_\delta)$ is scaled by a direction-dependent response $\text{sinc}\left[\frac{L}{\lambda_l}(\cos\theta_s - \cos\theta_l)\right]$. Therefore, the normalized beamforming pattern at this intermediate frequency is derived as

$$F_{\text{sp}}(\theta_s) = \frac{\Delta P(t)^2}{\max \Delta P(t)^2} = \text{sinc}^2\left[\frac{L}{\lambda_l}(\cos\theta_s - \cos\theta_l)\right]. \qquad (13)$$

**Derivation of multipeak beamforming.** We now derive the multipeak beamforming phenomenon of a quantum aperture dressed by multiple LO sources. With the composite LO field, the LO Rabi frequency becomes position dependent, $\Omega_l(y) = \frac{\mu_{34}}{\hbar}E_l(y) = \frac{\mu_{34}}{\hbar}\left|\sum_{n=1}^N E_{l,n} e^{jk_l y \cos\theta_{l,n} + j\phi_{l,n}}\right|$, and the LO-SIG phase difference function becomes

$$S(y,t) = \angle E_s(y,t) - \angle E_l(y,t) = \omega_s t + k_s y \cos\theta_s + \phi_s - \phi_l(y)$$
$$= \omega_\delta t + k_s y \cos\theta_s + \phi_s - \angle\left(\sum_{n=1}^N E_{l,n} e^{jk_l y \cos\theta_{l,n} + j\phi_{l,n}}\right). \qquad (14)$$

Therefore, substituting $\Omega_l(y)$ and $S(y,t)$ into Eq. (10) gives the AC laser transmission under multi-LO dressing

$$\Delta P(t) = P_{\text{in}} e^{-k_p \int_0^L \text{Im}[\chi_0(y)]dy} k_p \int_0^L \frac{\partial \text{Im}[\chi]}{\partial \Omega_l(y)} \Omega_s \cos(\omega_\delta t + k_s y \cos\theta_s + \phi_s - \phi_l(y)) dy,$$

$$= P_{\text{in}} e^{-k_p \int_0^L \text{Im}[\chi_0(y)]dy} k_p \text{Re}\left\{\left(\int_0^L \frac{\partial \text{Im}[\chi]}{\partial \Omega_l(y)} e^{j[k_s y \cos\theta_s - \phi_l(y)]} dy\right) \Omega_s e^{j(\omega_\delta t + \phi_s)}\right\}. \quad (15)$$

Unlike the single-LO case, the derivative term $\frac{\partial \text{Im}[\chi]}{\partial \Omega_l(y)}$ should remain inside the integral because the Rabi frequency of composite LO field $\Omega_l(y)$ varies along $y$. Eq. (15) shows that time-oscillating signal current $\Omega_s e^{j(\omega_\delta t + \phi_s)}$ is scaled by a direction-dependent response $\int_0^L \frac{\partial \text{Im}[\chi]}{\partial \Omega_l(y)} e^{k_s y \cos\theta_s - \phi_l(y)} dy$. This response gives a multipeak beamforming pattern

$$F_{\text{mp}}(\theta_s) \propto \left|\frac{\partial \text{Im}[\chi]}{\partial \Omega_l(y)} e^{j[k_s y \cos\theta_s - \phi_l(y)]} dy\right|^2 \propto \left|\frac{\partial \text{Im}(\bar{\rho}_{21})}{\partial E_l(y)} e^{j[k_s y \cos\theta_s - \phi_l(y)]} dy\right|^2. \quad (16)$$

## Data availability

The authors declare that the data supporting the findings of this study are available within the article and its Supplementary Information file. All other data that support the findings of the study are available from the corresponding authors upon request.

## Code availability

MATLAB (R2025b) codes used for processing data that support the findings of this study are available from the corresponding authors upon request.

## Author contributions

K. Huang and D. Zheng conceived the idea for the study. M. Cui derived the theoretical formula. M. Cui, Q. Zeng, and M. Chen designed the experiments and arranged specific tasks. M. Chen and Y. Wang built the platform. M. Cui, Q. Zeng, and M. Chen performed the experimental measurements with the assistance from Z. Zhu. M. Cui analysed the data. M. Cui and Q. Zeng wrote the manuscript and supplementary information with the assistance from M. Chen. The research was supervised by K. Huang, T. Mao, D. Zheng, and J. Zhang. All authors contributed to discussions regarding the results and analysis contained in the manuscript.

## Competing interests

The authors declare no competing financial interests.

## Additional information

Supplementary Information accompanies this paper.

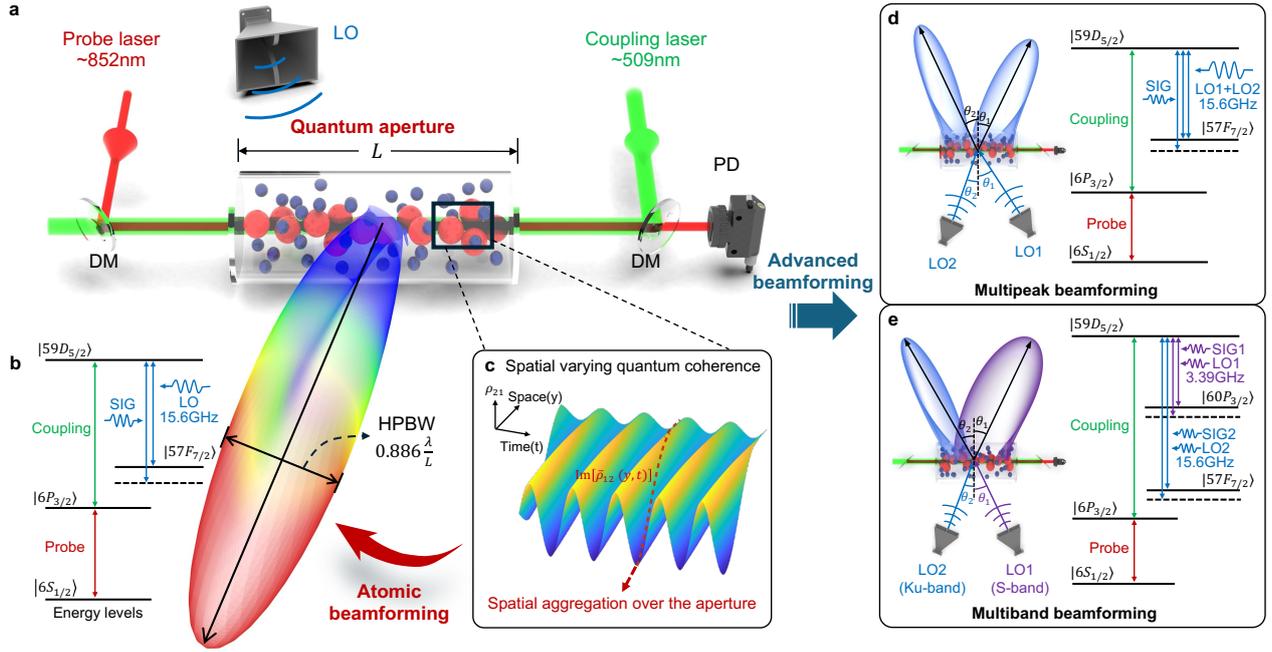

**Fig. 1 | Operating principle of quantum-aperture beamforming.** We have used the following notations: (1) PD: photodetector, (2) LO: local oscillator, (3) SIG: signal, (4) HPBW: half-power beamwidth, (5) DM: dichroic mirror. **a**, Schematics of quantum-aperture beamforming system. A vapor cell of length $L$ contains Rydberg atoms prepared by counter-propagating probe and coupling lasers. An LO source illuminates the vapor cell from the spatial direction $\theta_l$ to enable detection of a SIG field arriving direction $\theta_s$. **b**, Energy-level diagram. Atoms are excited from the ground state $|6S_{1/2}\rangle$ to the Rydberg state $|59D_{5/2}\rangle$ by a probe laser (852 nm) and a coupling laser (509 nm). The LO/SIG fields couple the transition $|59D_{5/2}\rangle \rightarrow |57F_{7/2}\rangle$. **c**, Interference between the LO and SIG fields produces a spatially varying quantum coherence inside the atom medium. This endows the quantum aperture with a direction-dependent response, which thereby produces a beam. The beam is steered to the direction opposite the LO, with an HPBW of $0.886\lambda/L$, where $\lambda$ is the wavelength. **d,e**, Advanced beamforming enabled by quantum aperture. (**d**) When multiple LO sources illuminate the vapor cell from different directions, the quantum aperture can generate multipeak beams. (**e**) Multiple LO sources, operating at different frequency bands, illuminate the vapor cell and trigger distinct Rydberg transitions (e.g., 3.39GHz couples to $|59D_{5/2}\rangle \rightarrow |60P_{3/2}\rangle$, and 15.59GHz couples to $|59D_{5/2}\rangle \rightarrow |57F_{7/2}\rangle$). This configuration enables simultaneous beamforming at multiple bands within a single vapor cell.

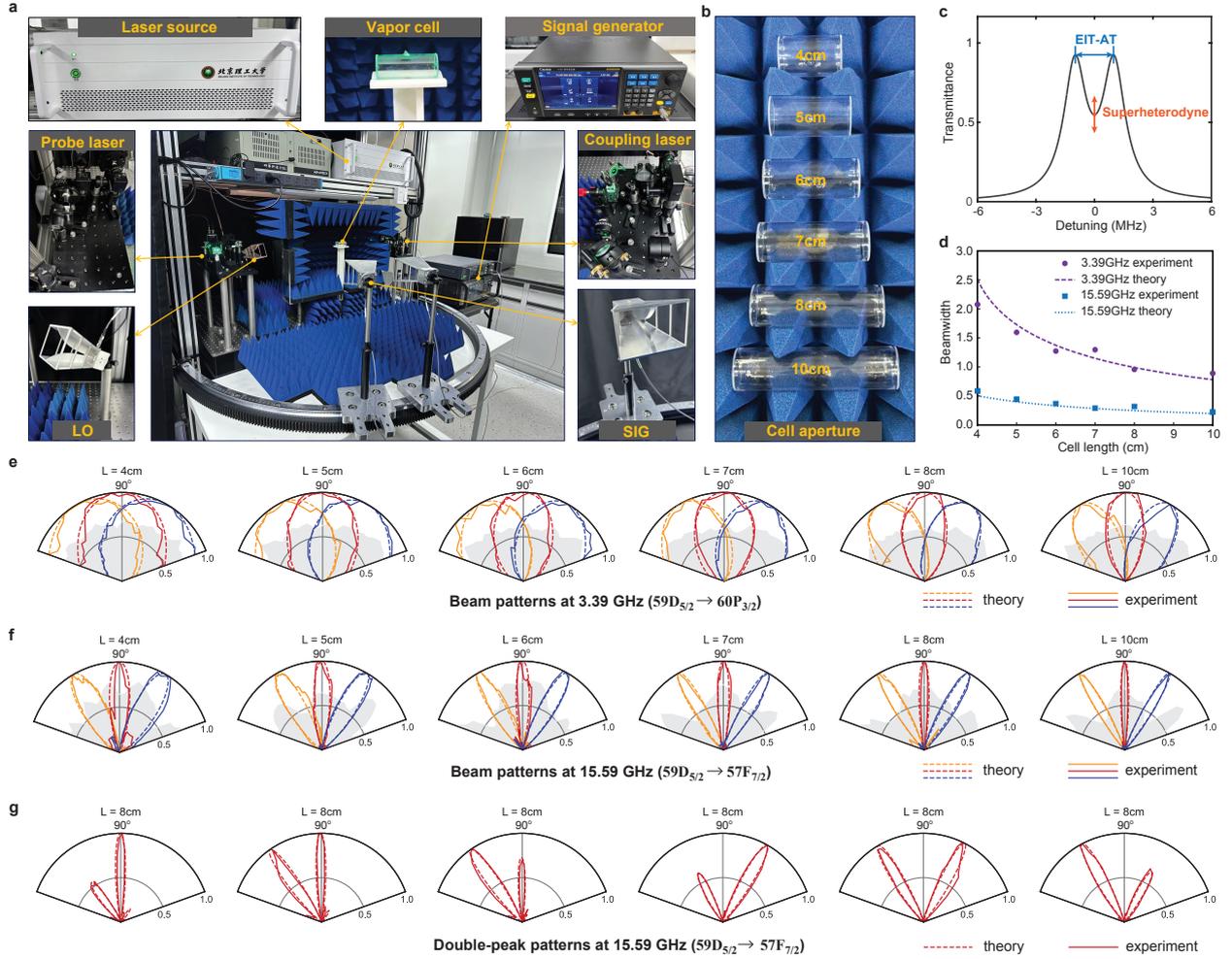

**Fig. 2 | Experimental demonstration of quantum-aperture beamforming. a**, Experimental platform comprising an atomic vapor cell, LO horn antennas, and SIG horn antennas mounted on a 1.4-m-diameter semicircular rail. The SIG source is rotated from 20° to 160° along the rail to measure the angular response (see details in Methods). **b,** Vapor cells with lengths ranging from 4cm to 10cm. **c,** Comparison of beam patterns obtained via EIT-AT detection (without an LO) and superheterodyne detection (with LOs). The former extracts the incident field strength from the EIT-AT peak interval, whereas the latter monitors the AC component of the transmitted probe laser. **d,** Influence of cell length and frequency band on the theoretical (lines) and experimentally measured (symbols) beamwidths for superheterodyne detection. **e,f,** Measured beam patterns for EIT-AT detection (shaded regions) and for superheterodyne detection with a single LO (solid lines), together with theoretical predictions of superheterodyne detection (dashed lines). Results are shown for six cell lengths: 4/5/6/7/8/10 cm, and three LO directions: 240° (yellow lines), 270° (red lines), 300° (blue lines). Beam patterns were measured for the $|59D_{5/2}\rangle \to |60P_{3/2}\rangle$ transition (**e**) at 3.39 GHz, and for the $|59D_{5/2}\rangle \to |57F_{7/2}\rangle$ transition (**f**) at 15.59 GHz. **g,** Double-peak beamforming results at 15.59 GHz with the cell length $L = 8$ cm. Measured patterns (solid lines) and theoretical fits (dashed lines) are shown for two LO configurations: LO directions (240°, 270°) for the left three panels; LO directions (240°, 300°) for the right three panels. In each panel, three LO-power pairs are tested: (3 dBm, 6 dBm), (4 dBm, 4 dBm), and (6 dBm, 3 dBm).

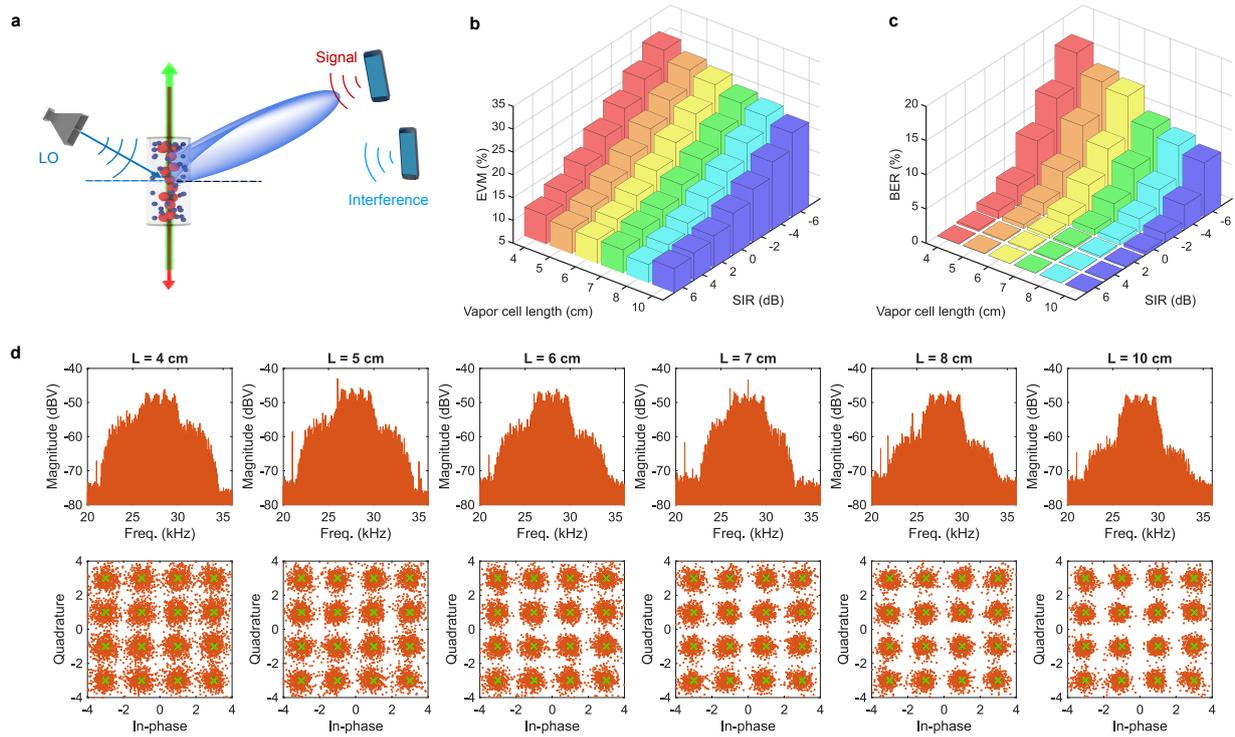

**Fig. 3 | Experimental results of quantum aperture enabled interference mitigation. a**, Schematic of a beamforming system supporting interference mitigation. In this demonstration, signal sources at 15.59 GHz illuminate the vapor cell from 60° and 75° to serve as a signal-of-interest and an interferer, respectively. The signal-of-interest transmits a 16-QAM data stream at a symbol rate of 4 kSymbols/s while the interferer emits random symbols with a bandwidth of 8 kHz. The receiver decodes the stream from signal-of-interest. **b,c,** Influence of the vapor cell length and transmit signal-to-interference ratio (SIR) on the system performance measured by: (**b**) error vector magnitude (EVM) and (**c**) bit error ratio (BER). **d**, Received spectra (intermediate frequency: 28 kHz) and demodulated 16-QAM constellation diagrams for varying vapor cell lengths from 4 cm to 10 cm. The spectra show a progressive increase in the received SIR, and the progressive concentration of constellation points corresponds to the reduced EVM and BER.

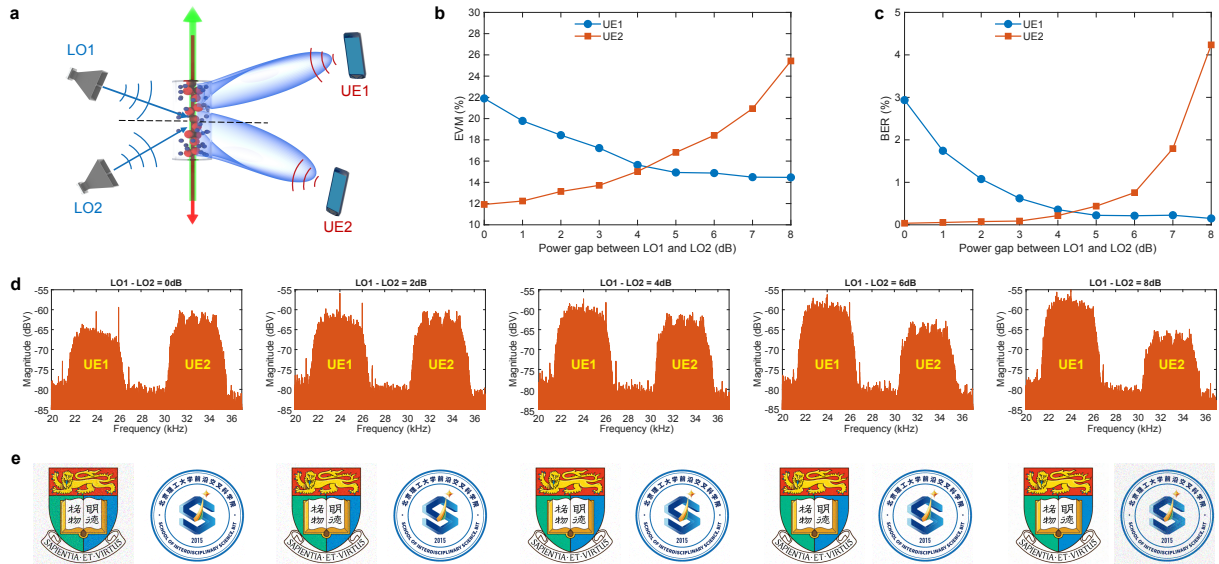

**Fig. 4 | Experimental results of quantum aperture enabled multiuser access. a**, Schematic of a beamforming system supporting uplink access of multiple devices operating in the same frequency band. In this demonstration, two LO sources at 15.59 GHz illuminate the vapor cell from 300° and 270°, respectively, to support radio access for two signal sources at 60° and 120°. Each source transmits a 16-QAM data stream at a symbol rate of 4 kSymbols/s, and the receiver decodes the two streams independently. **b,c,** Influence of the two LOs' power on the system performance measured by (**b**) EVMs and (**c**) BERs regarding the two users (UE1 and UE2). The power of LO1 increases from 6 dBm to 10 dBm, while the power of LO2 decreases from 6 dBm to 2 dBm. **d**, Received spectra (intermediate frequencies: 24 kHz for UE1, 33 kHz for UE2) for varying LOs' power, with power gap between LO1 and LO2 increasing from 0 dB to 8 dB. **e**, System performance illustrated by uncoded transmission of university logos (UE1 left, UE2 right).

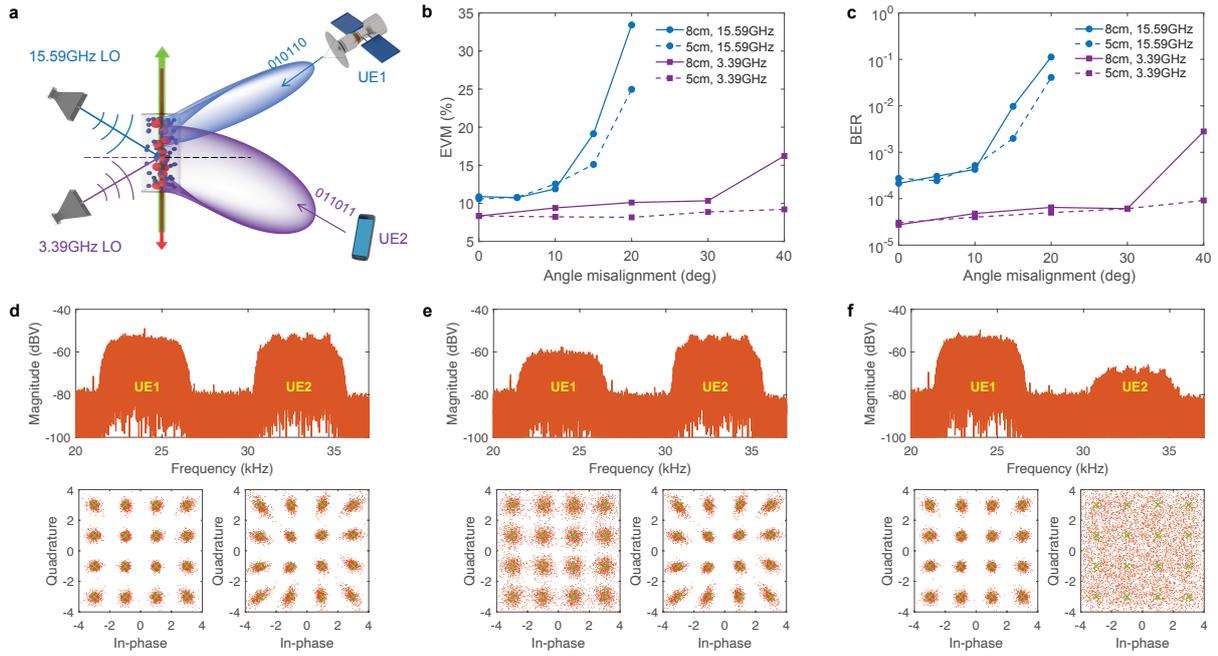

**Fig. 5 | Experimental results of quantum aperture enabled multiband multiuser access. a**, Schematic of a multiband beamforming system supporting simultaneous uplink access for heterogeneous devices operating in different frequency bands. LO sources at 3.39 GHz and 15.59 GHz illuminate the vapor cell from 300° and 240°, respectively, to detect an S-band SIG source at 60° and a Ku-band SIG source at 90°. Each SIG source transmits a 16-QAM data stream with a rate of 4 kSymbols/s, while the receiver decodes the two streams independently. **b,c,** Influence of angle misalignment on the achieved (**b**) EVMs and (**c**) BERs of the multiband multiuser access system. Communication performance for both 5-cm and 8-cm vapor cell were measured. The S-band source was misaligned from its beam center by 0°-40° and the Ku-band source by 0°-20°. Performance degrades more sharply for the Ku-band due to its narrower beamwidth. **d-f,** Received spectra (intermediate frequencies: 24 kHz for S-band, 33 kHz for Ku-band) and recovered 16-QAM constellations under three alignment conditions: (**d**) both signal sources perfectly aligned; (**e**) S-band source misaligned by 40° (Ku-band aligned); (**f**) Ku-band source misaligned by 20° (S-band aligned).